# Coherent stacking of laser pulses in a high-Q optical cavity for accelerator applications*

*V.P. Androsov, I.M. Karnaukhov, Yu.N. Telegin*

*National Science Center "Kharkov Institute of Physics and Technology", Kharkov, Ukraine*
e-mail: androsov@kipt.kharkov.ua

## Abstract.

We have performed the harmonic analysis of the steady-state coherent pulse-stacking process in a high-Q Fabry-Perot cavity. The expression for the stacked pulse shape is obtained as a function of both the laser cavity and pulse-stacking cavity parameters. We have also estimated the pulse power gains attainable in the laser-optical system of NESTOR storage ring, which is under development at Kharkov Institute of Physics and Technology. It is shown that high power gains ($\sim 10^4$) can be, in principle, achieved in a cavity, formed with low-absorption, high reflectivity ($R \approx 0.9999$) mirrors, if the laser carrier frequency will be matched to the second harmonic frequency of the pulse-stacking cavity. This means a development of the sophisticated frequency stabilization loop for maintaining the cavity length constant within a sub-nanometer range.

*PAC numbers: 29.20 Dh, 42.60By, 42.60 Da*

## Introduction

Over the last two decades Compton scattering (CS) of the laser light by the electrons circulating in the storage ring was extensively used for production of the quasi-monochromatic highly polarized γ-beams in the energy range from 30 MeV to 2 GeV for nuclear physics experiments [1]. Implementation of this technique at medium- and low-energy circular machines was hampered by the strong space-charge effects. After publication of Telnov's article on laser-electron cooling [2] a number of projects has appeared where CS has been considered as a tool for production of the intensive X-ray beams at low-energy ($E_0$=10÷200 MeV) compact electron storage rings [3,4]. Lastly, the most promising and, at the same time, the most challenging are the schemes of using CS for e-γ - conversion in linear supercolliders [5,6].

The common feature of all these facilities is the problem of obtaining the intensive laser pulses with pulse width of ~10 ps and pulse energy of ~10 mJ (for γγ - colliders these values are ~1 ps and ~1 J, respectively) that follow at a repetition frequency of 100÷500 MHz. The only feasible solution that naturally suggests itself is coherent stacking of laser pulses with required time characteristics in a high-Q optical cavity system. The latter can be a simple two-mirror open cavity (Fabri-Perot cavity) [7,8], a two-mirror cavity with one compound mirror for extraction of the secondary photon beam [9], or it can represent a pair of optical cavities, one of them imbedded into the other [10,11]. In the last case the external cavity can comprise a low-gain regenerative amplifier for compensation of energy losses in the system [10].

Today the stacking of continuous wave (CW) laser beams in a Fabry-Perot cavity is a well-established technique, and it is extensively used in various fields of physics: gravitational interferometers [12], laser-wire beam profile monitors [13], etc. Implementation of this technique for quasi-CW laser beams (henceforth, we apply this term for a continuous succession of short laser pulses following at a high repetition frequency from a mode-locked laser) is encountered with difficulties. The essential requirement to the resonance optical system, intended for stacking of short laser pulses, is coherence of pulse-summing process that ensures accumulation of energy in the laser pulse without deterioration of its spatial and temporal characteristics. To meet this requirement the axial-mode spectrum of the pulse-stacking cavity (PSC) has to match the harmonic spectrum of quasi-CW laser beam, the latter being coincident with the mode spectrum of the laser cavity with active element.

---

*Work supported by NATO under the project sfp-977982

Feasibility of coherent pulse stacking in a high-Q open cavity was recently demonstrated by R.J. Loewen at SLAC [8]. Iteratively adjusting the cavity length eventually enabled the laser to lock in 2÷3 ms intervals to the peak axial mode in a 6.7 kHz bandwidth cavity (mirrors reflectivity $R$=0.9998). The accumulation factor was estimated to be ≈4500. It should be noted that the natural, manufacturer-specified, laser pulse width of 7ps was stretched up to 25-30 ps in order to eliminate the significant dispersion effects. The effect of dispersion mismatch between the laser cavity and PSC was also discussed in this work and assumed to be insignificant for pulse widths considered. To evaluate dispersion effects correctly one has to solve the problem in the frequency domain, i.e. perform a harmonic analysis of the stored pulse.

In this paper we present the results of sequential harmonic analysis of the steady-state pulse-stacking process in a high-Q Fabry-Perot cavity together with some estimations for NESTOR storage ring, which is under development at Kharkov Institute of Physics and Technology [4].

## *1. General description.*

In general, it is rather difficult to gain matching between the laser and the PSC, because not only cavity lengths have to be matched but also their frequency spectra have to be identical. The main causes that hamper the matching are dispersion in the optical elements and the essential difference in reflector parameters that form these cavities. The last ensues from different requirements set up to the laser and pulse-stacking cavities.

The laser cavity reflectors have a large curvature radius in order to form a wide beam thus ensuring effective interaction with the lasing medium all around the cavity. Reflectivity of the mirror, through which the laser beam is extracted, has to be low ($R$~0.9) in order to attain a reasonable efficiency of the laser system and to provide the required output power.

The mirrors of the PSC have to meet practically contrary requirements. For achievement of high accumulation factors ($k_{ph}$=$10^3$÷$10^4$) they have to be high-reflectivity mirrors ($R$=0.999÷0.9999). The radius of curvature $\rho_c$ has to be rather small ($\rho_c$≈$L_c$/2, where $L_c$ is cavity length) in order to focus the laser beam to the required spot size in the point of its interaction with the electron beam.

The axial mode spectrum $f_q$ of the symmetric cavity, formed with two mirrors with complex reflectivity $\dot{r}_c = r_c \exp(-i\varphi_{\dot{r}_c})$, is given by [11]:

$$f_q = f_{FSR}\left[q+1+\frac{2}{\pi}\arctan\left(\frac{\rho_c}{L_c/2}-1\right)^{-1/2} - \varphi_{\dot{r}_c}/\pi\right], \qquad (1)$$

where $q$=2$L_c$/$\lambda$ is the longitudinal index of the axial cavity mode $TEM_{00q}$; $\lambda$ is the laser wavelength; $f_{FSR}$=$c$/2$L_c$ is the free spectral range and $c$ is the speed of light. The term $\frac{2}{\pi}arctg\left(\frac{\rho_c}{L_c/2}-1\right)^{-1/2}$ gives the mode frequency shift due to the wave-front curvature tied in with sphericity of the cavity mirrors. For the plane-parallel cavity ($\rho_c$→∞) this term vanishes while for the concentric cavity it tends to unity. For these two extreme cases the difference in mode frequencies with the same longitudinal index $q$ is equal to free spectral range $f_{FSR}$. The similar effect on the spectra of cavity modes produces a difference in the phases $\varphi_{\dot{r}_c}$ of complex reflectivity of the cavity mirrors. For the real parameters of the laser and pulse-stacking cavities their spectra are shifted one against the other by the value less than $f_{FSR}$.

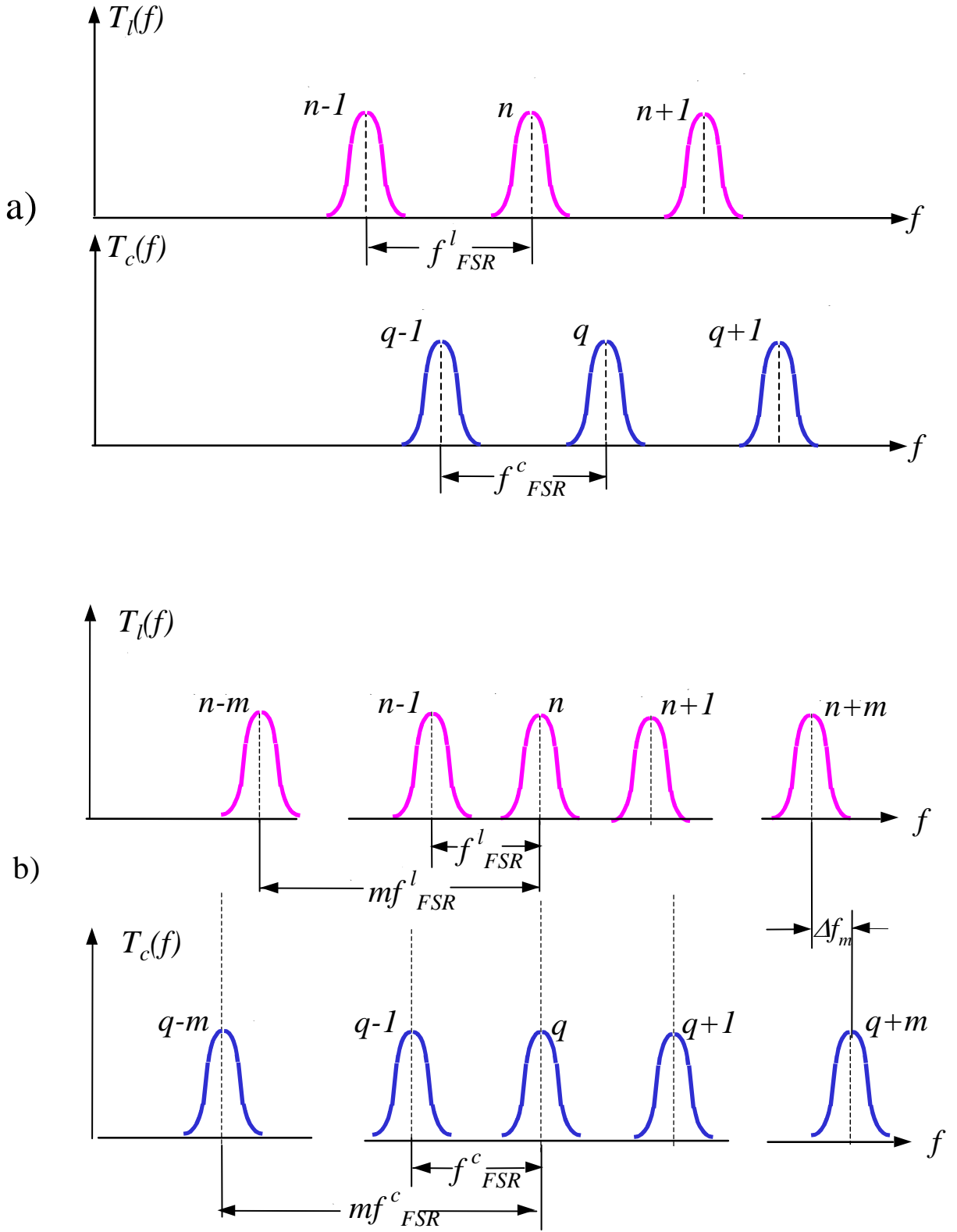

Fig. 1. Schematic spectra of the laser cavity modes and PSC modes: a) $L_l = L_c$, no mode matching; b) after adjusting PSC length ($L_l \neq L_c$) for mode matching ($f^l_n = f^c_q$).

This effect is illustrated in Fig. 1a where the spectra of two cavities, namely, laser cavity ($L_l$, $\rho_l$, $\varphi_{\dot{r}_l}$, $n = 2L_l/\lambda$) and PSC ($L_c$, $\rho_c$, $\varphi_{\dot{r}_c}$, $q$) are sketched. Here the laser cavity with active element is presented with some equivalent two-mirror cavity because, as follows from ref. [8], we can neglect the dispersion in optical elements of the picosecond laser. The cavity lengths are matched, i.e. $L_l = L_c$. To drive the PSC at the laser carrier frequency $\omega_0 = 2\pi f_0 = 2\pi c/\lambda$, which corresponds to the fundamental cavity mode with longitudinal index $n$, one has to match this frequency to the frequency of the PSC fundamental mode with longitudinal index $q$ or to the frequency of any of PSC modes lying in close proximity to the fundamental one. It can be done either by changing the carrier frequency via a proper adjustment of the laser cavity length or by changing the PSC length. In both these cases the free spectral ranges for two cavities become unequal ($f^l_{FSR} \neq f^c_{FSR}$), so the mode frequencies in two cavities get different shifts against the matched mode frequency $f^l_n = f^c_q$. Fig. 1b illustrates how the frequencies of the PSC modes shift in both sides from the matched-mode frequency when the PSC length is adjusted for mode matching. The frequency shift between two modes, equidistant from the matched modes by number $m$, can be obtained from the following relation:

$$\Delta f_m = f^l_{n+m} - f^c_{q+m} = -m f^c_{FSR} \cdot \frac{\Delta L}{L_l}, \qquad (2)$$

where $\Delta L = L_l - L_c$. One can see that this shift is proportional to the relative difference of the laser cavity and PSC lengths, and it linearly increases for equidistant harmonics with their displacement from the matched modes. So, the quasi-CW laser beam sidebands will drive the PSC at frequencies that correspond to the wings of its resonance curves instead of their peaks. It changes the phases and amplitudes of harmonic components of the stored pulse against their values in the incident laser pulse. It can lead to distortion of the time profile of the stored picosecond laser pulse and to it's lengthening that finally results in lower accumulation factors achieved.

Below we present the results of study of pulse-stacking process in the high-Q two-mirror cavity by methods of harmonic spectral analysis. Absorption in cavity mirrors is not taken into account (A=0), because our preliminary study shows that it does not affect the time characteristics of the stored pulse while essentially complicates the derived expressions. Absorption reduces the accumulation factor (especially, for ultra high-reflectivity mirrors $R \sim 0.9999$) and can be easily taken into account as a correction to the formulae obtained in this work.

## *2. Harmonic analysis of the pulse-stacking cavity.*

The infinite periodical sequence of electromagnetic pulses of arbitrary shape propagating in z-direction with the group velocity $v$ and the repetition time $T_{rep} = 1/f_{rep}$ can be presented in the time domain with the infinite sum of pulses:

$$S(t,z) = \sum_{n=-\infty}^{\infty} \int_{-\infty}^{\infty} s(\tau) \cdot \delta\left[\tau - \left(t + nT_{rep} - \frac{z - z_0}{v}\right)\right] d\tau, \qquad (3)$$

each of them presented with the function $s(\tau) = f(\tau) Cos(\omega_0 \tau + \delta_0)$, where $\omega_0$ and $\delta_0$ are carrier frequency and carrier initial phase, respectively. The pulse structure is displayed in Fig. 2.

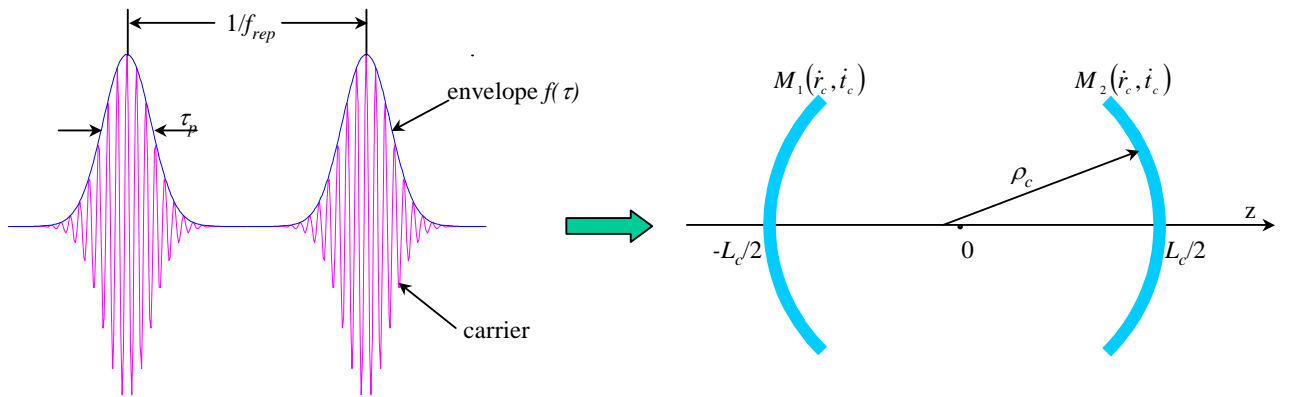

Figure 2. The laser pulse structure and the PSC parameters

Henceforth, we assume the Gaussian form of the laser pulse envelope $f(\tau) = \exp[-2(t/\tau_p)^2]$, where $\tau_p$ is the pulse duration. Let then assume that this sequence of pulses is incident on the symmetrical two-mirror PSC with complex mirror reflectivity $\dot{r}_c$ and complex transmission factor $\dot{t}_c$. Note that $R=\dot{r}_c^2$, $T=\dot{t}_c^2$ and $R+T=1$.

Let next proceed to the harmonic analysis, which is carried out in several steps.

## 2.1 Direct Fourier transform of the incident laser beam.

The incident laser beam has to be presented in the frequency domain at some fixed plane. In our case it is the plane $z=-L_c/2$ where the left mirror of the PSC is located:

$$s(t,-L_c/2) = \frac{1}{2}\left\{\sum_{m=-\infty}^{\infty}\dot{C}_m \exp\langle i[(\omega_0 + m\omega_{rep})t + \delta_0]\rangle + \sum_{m=-\infty}^{\infty}\dot{C}_m \exp\langle -i[(\omega_0 - m\omega_{rep})t + \delta_0]\rangle\right\}, \quad (4)$$

where $\omega_{rep}=2\pi/T_{rep}$; ; and Fourier coefficients $\dot{C}_m$ is given by:

$$\dot{C}_m = \frac{1}{T_{rep}}\int_{-T_{rep}/2}^{T_{rep}/2} f(\tau)\exp(-im\omega_{rep}t)d\tau, \quad (5)$$

For the Gaussian pulses $\dot{C}_m$ can be obtained with the following relation:

$$\dot{C}_m = \sqrt{\frac{\pi}{2}} \cdot \frac{\tau_p}{T_{rep}}\exp\left[-\left(m\pi\frac{\tau_p}{T_{rep}}\right)^2/2\right] \quad (6)$$

## 2.2. The PSC eigenwaves

The PSC eigenwaves propagating along z-axis can be presented with two sets of equations:

$$E_m^+ \sim \exp\left\langle -i\left\{k_m^+(z+L_c/2) - \arctan\left(\frac{z}{z_{Rm}^+}\right) + \arctan\left(\frac{-L_c/2}{z_{Rm}^+}\right) - [(\omega_0 + m\omega_{rep})t + \delta_0]\right\}\right\rangle, \quad (7a)$$

$$E_m^- \sim \exp\left\langle i\left\{k_m^-(z+L_c/2) - \arctan\left(\frac{z}{z_{Rm}^-}\right) + \arctan\left(\frac{-L_c/2}{z_{Rm}^-}\right) - [(\omega_0 - m\omega_{rep})t + \delta_0]\right\}\right\rangle, \quad (7b)$$

where: $k_m^\pm = \frac{\omega_0 \pm m\omega_{rep}}{c} = \frac{2\pi}{\lambda_m^\pm}$; $\lambda_m^\pm$ is the wavelength of the harmonic with a harmonic number $\pm m$; $z_{Rm}^\pm = \pi w_0^2/\lambda_m^\pm$ are Rayleigh lengths for the corresponding harmonics; $2w_0$ is the transverse beam waist size at the center of cavity ($z=0$). The eigenwaves $E_m^+$ correspond to the harmonics with time dependence of $\exp\langle i[(\omega_0 + m\omega_{rep})t + \delta_0]\rangle$, while $E_m^-$ correspond to the harmonics with time dependence of $\exp\langle -i[(\omega_0 + m\omega_{rep})t + \delta_0]\rangle$

Considering that curvature of the wavefronts for all harmonics in the PSC is defined with the same reflectors their Rayleigh lengths $z_{Rm}^\pm$ are equal and the uniform Rayleigh length $z_R$ can be obtained with the following relation:

$$\frac{z_R}{L_c/2} = \left(\frac{\rho_c}{L_c/2} - 1\right)^{1/2} \quad (8)$$

## 2.3. Determination of the transfer functions

Assuming the PSC is dispersion-free, it is sufficient to derive the steady-state transmission factors through the mirror $M_1$ ($z=-L_c/2$) for each harmonic in order to obtain the parameters of the laser pulse stored in the PSC. The strict and detailed consideration for two sets of eigenwaves (7a, 7b) shows that the transfer function for $E_m^+$ is given by:

$$\dot{T}_m^+ = \frac{\dot{t}_c}{1 - \dot{r}_c^2 \cdot \exp\left[-2i\left(k_m^+ L_c - 2\arctan\frac{L_c/2}{z_R}\right)\right]} \qquad (9)$$

The transfer function for $E_m^-$ (for some reasons, that ensue from the form of $E_m^+$ and $E_m^-$, we denote it as $(\dot{T}_m^-)^*$) can be found as complex conjugate of the right-hand side of Eq. (9) in which $m$ is substituted with $-m$:

$$(\dot{T}_m^-)^* = \left\{\frac{\dot{t}_c}{1 - \dot{r}_c^2 \cdot \exp\left[-2i\left(k_m^- L_c - 2\arctan\frac{L_c/2}{z_R}\right)\right]}\right\}^* \qquad (10)$$

It also follows from the simple physical considerations: the phase shift along some fixed path does not depend on complex representation of the eigenwave.

## 2.4. Solution for the stored beam

The beam stored in the PSC can be presented at $M_1$ location with the following infinite Fourier series:

$$B(t, -L_c/2) = \frac{1}{2}\left\langle \sum_{m=-\infty}^{\infty} \dot{C}_m \dot{T}_m^+ \exp\{i[(\omega_0 + m\omega_{rep})t + \delta_0]\} + \sum_{m=-\infty}^{\infty} \dot{C}_m (\dot{T}_m^-)^* \exp\{-i[(\omega_0 - m\omega_{rep})t + \delta_0]\}\right\rangle \qquad (11)$$

From Eq. (6) follows that $\dot{C}_{-m} = \dot{C}_m$. From a comparison of Eqs (9) and (10) ensues that $\dot{T}_{-m}^- = \dot{T}_m^+$ and $\dot{T}_m^- = \dot{T}_{-m}^+$, so Eq. (11) can be reduced to:

$$B(t, -L_c/2) = \sum_{m=-\infty}^{\infty} \dot{C}_m T_m^+ \cos[(\omega_0 + m\omega_{rep})t + \delta_0 + \Psi_m^+], \qquad (12)$$

where $T_m^+$ and $\Psi_m^+$ are, respectively, the modulus and the phase of the transfer function $\dot{T}_m^+ = T_m^+ \cdot \exp(i\Psi_m^+)$. They can be obtained from the following relations:

$$T_m^+ = \frac{t_c}{1 - r_c^2} \frac{1}{\sqrt{1 + \left(\frac{2r_c \sin\alpha_m^+}{1 - r_c^2}\right)^2}}, \qquad (13)$$

$$\Psi_m^+ = -\arctan\frac{r_c^2 \sin 2\alpha_m^+}{1 - r_c^2 \cos 2\alpha_m^+} - \varphi_c^i, \qquad (14)$$

where:

$$\alpha_m^+ = k_m^+ L_c - 2\arctan\left(\frac{\rho_c}{L_c/2} - 1\right)^{-1/2} + \varphi_c^r, \qquad (15)$$

and $\varphi_c^i$, $\varphi_c^r$ are the phases of the transmission factor $\dot{t}_c$ and reflectivity factor $\dot{r}_c$, respectively. Considering that the laser carrier frequency has to be resonant for the PSC one can reduce expressions (13), (14) to:

$$T_m^+ = \frac{k_{ph}^0}{\sqrt{1 + \left[\frac{2r_c \sin(m\pi \Delta L/L_l)}{1-r_c^2}\right]^2}} \tag{16}$$

$$\Psi_m^+ = \arctan \frac{r_c^2 \sin(2m\pi \Delta L/L_l)}{1 - r_c^2 \cos 2m\pi \Delta L/L_l} - \varphi_c^t \tag{17}$$

where: $\Delta L/L_l = (L_l - L_c)/L_l$, and $k_{ph}^0$ is the amplitude gain in the PSC for the first harmonic of the laser beam given by:

$$k_{ph}^0 = t_c/(1 - r_c^2) \tag{18}$$

One can derive the expression for $\Delta L/L_l$ by using the dispersion relations for the PSC and the equivalent laser cavity at the carrier frequency:

$$k_0 L_c - 2\arctan\left(\frac{\rho_c}{L_c/2} - 1\right)^{-1/2} + \varphi_c^r = \pi(q+1) \tag{19a}$$

$$k_0 L_l - 2\arctan\left(\frac{\rho_l}{L_l/2} - 1\right)^{-1/2} + \varphi_l^r = \pi(n+1) \tag{19b}$$

where indices $c$ and $l$ refer to the pulse-stacking and laser cavities, respectively. The result is given by:

$$\Delta L/L_l = 1 - \frac{q}{n} + \frac{\left\{\left[2\arctan\left(\frac{\rho_l}{L_l/2} - 1\right)^{-1/2} - \arctan\left(\frac{\rho_c}{L_c/2} - 1\right)^{-1/2}\right] - (\varphi_l^r - \varphi_c^r)\right\}/\pi}{n} \tag{20}$$

By substituting Eqs (16), (17) and (20) in Eq. (12) one can obtain the final solution for the stored pulse at $z=-L_c/2$.

## *3. Calculation results and discussion*

Before performing calculations let's look closely at the derived formulae in order to analyze the effect of parameters of the laser-optical system on the stored pulse characteristics.

The expressions (16), (17) describe a behavior of the modulus and phase of the PSC transfer function for all laser beam harmonics. It is seen that in the case of the ideal matching of the PSC and laser beam spectra ($\Delta L/L_l=0$) all harmonics do not sustain any changes except the same amplitude enhancement by factor $k_{ph}^0$, so the stored pulse keeps the original shape of the incident laser pulse. By the way, factor $k_{ph}^0$ gives the maximal value of pulse amplitude gain, which can be achieved in two-mirror stacking cavity with non-absorptive mirrors. If $\Delta L/L_l \neq 0$ the modules and phases of the transfer functions for various harmonics differ, and this difference increases as the frequency shift between harmonic and carrier frequency increases. Since the distant harmonics define mainly the pulse fronts, one can anticipate a widening of the stored pulse (in the time domain) against the incident one. For the shorter pulses this effect will be more pronounced than for the longer ones. Another important parameter strongly affecting the stored pulse shape is the reflectivity of the PSC mirrors $r_c$.

We performed calculations by using the derived formulae in order to evaluate the influence of the PSC parameters and the incident pulse width on the stored pulse characteristics. The calculations were made for the parameters that are relevant to the laser-optical system of NESTOR facility [4], where we are intended to use the mode-locked Nd:YAG laser ($\lambda$=1064 μm) with an average beam power of 10 W. Two possible PSC configurations were considered: 0.42m (short) cavity and 2.52m (long) cavity. The first one corresponds to facility operation with 18 electron bunches (every second electron bunch of the total 36 is used for interaction) and a high laser repetition rate $f_{rep}$=350 MHz. The long cavity is now considered as a backup version for operation with only 3 bunches and a relatively low repetition rate of laser pulses ($f_{rep}$≈58 MHz). The first variant is preferable in aspect of efficiency of using the electron beam, because no sub-harmonic prebunching of the injected beam is provisioned for NESTOR, and all RF-buckets will be filled.

The radius of curvature of PSC mirrors $\rho_c$≈21.4cm was chosen so as to obtain the transverse beam-waist size $2w_0$ =200μm. The same parameter for laser mirrors was varied so as to obtain the beam-waist size in the laser cavity 1mm and 2mm. The reflectivity of the laser mirrors was taken to be 0.95, while PSC mirrors reflectivity ranged from 0.999 to 0.9999.

In Fig. 3 the time profile of the stored pulse is presented both for the short (a, c) and long (b, d) cavity and for medium-reflectivity (a, b) and high-reflectivity (c, d) mirrors. The incident laser pulse width 7ps (FWHM) corresponds to the specified value of picoTRAIN series of mode-locked lasers manufactured by High-Q Laser Production GmbH [14]. In vertical scale the instant power in stored laser pulse - $I_{stored}$ is plotted in arbitrary units. The incident laser pulse amplitude is unity, so from a comparison of the peak power value from the figure with the square of $k_{ph}^0$, calculated by using the simple relation (18), we can deduce how close to the ideal matching we approach. Note, that $(k_{ph}^0)^2$ is equal to $10^3$ and $10^4$ for (a, b) and (c, d), respectively.

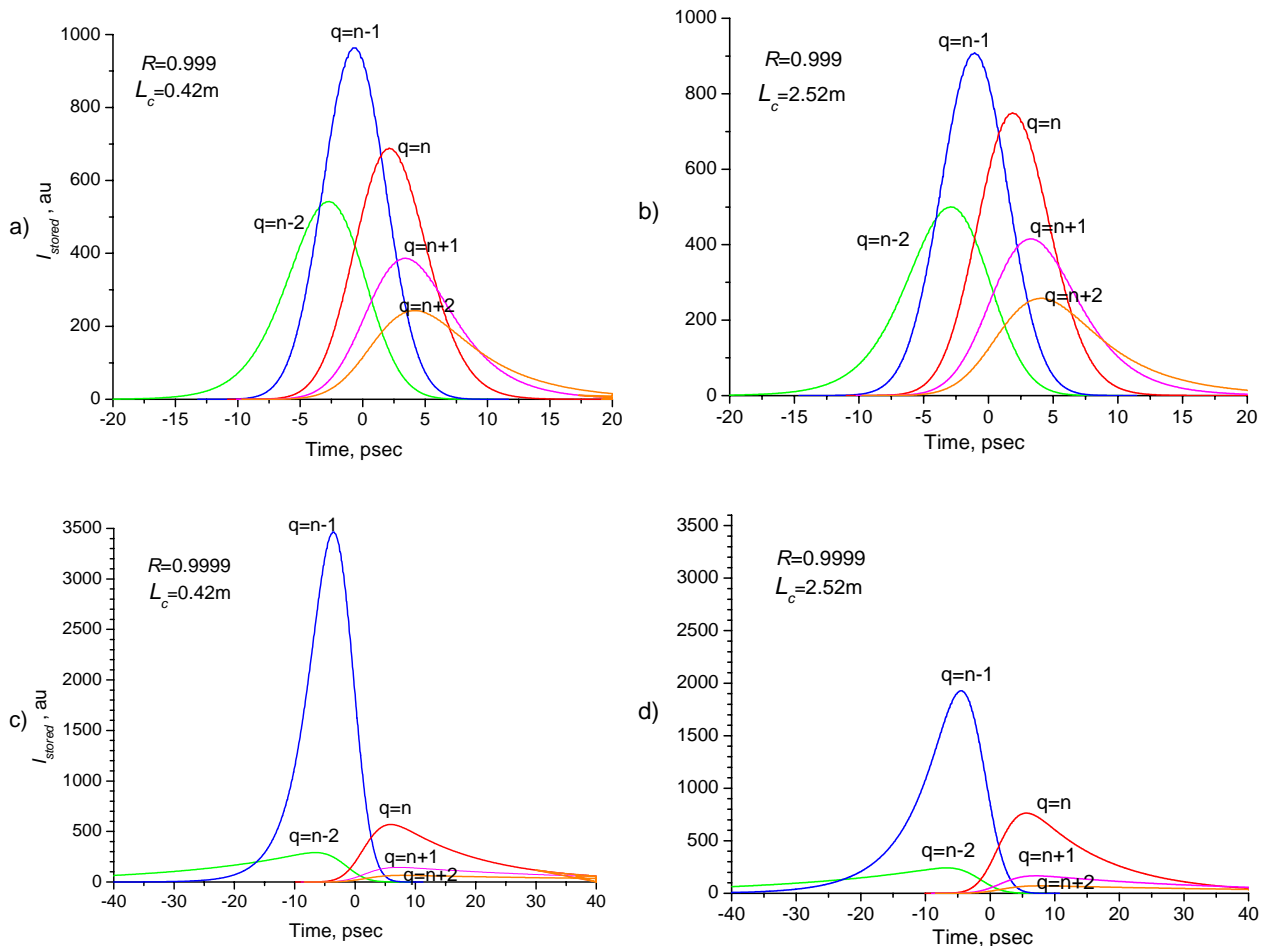

Fig. 3. Time profiles of the 7 ps laser pulse stored in the short (a, c) and long (b, d) PSC with medium-reflectivity (a, b) and high-reflectivity (c, d) mirrors. The family of curves in each picture corresponds to different shifts between longitudinal indices of the matched laser and PSC modes.

One can see that in all pictures the best results are obtained for the case, when the PSC axial mode $TM_{00q}$ is matched to the laser cavity mode $TM_{00n-1}$. We can also deduce from the figure the following conclusions:

- for medium-reflectivity mirrors we come very close to the ideal matching results, i.e. $(k_{ph}^0)^2 \sim 10^3$, while for the PSC with high-reflectivity mirrors we are far from the goal value of $10^4$;
- for the PSC with high-reflectivity mirrors only one cavity mode can provide the reasonable (for given mirror parameters) amplitude gain; the cavity with medium-reflectivity mirrors permits one to use several modes, thus simplifying mode matching and widening the operation range;
- the long-cavity version yields to the short-cavity one, this disadvantage becomes more pronounced with an increasing of mirror reflectivity;
- pulse-shape distortion and pulse widening is conspicuous only for the non-optimal mode-matching, except the case of long PSC with high reflectivity mirrors; more detailed consideration shows that this effect is more noticeable for the high-reflectivity mirrors.

The last effect is seen clearly in Fig. 4 where the shape of the stored pulse is given for different reflectivity of the PSC mirrors and two different radii of curvature of the laser mirrors. Only the optimal inter cavity mode shift $q=n\text{-}1$ is presented. Solid lines correspond to large curvature radius of the laser mirror $\rho_l=41.7$ m providing 2mm beam waist in the laser crystal, while the dashed lines correspond to $\rho_l=3.2$ m that provides 1mm beam waist. In the last case the essential decreasing of the stored pulse power gain is seen for high-reflectivity mirrors. The noticeable widening of the stored pulse for the PSC with high-reflectivity mirrors is also seen: by factor 1.5 and 1.8 for $\rho_l=41.7$ m and $\rho_l=3.2$ m, respectively.

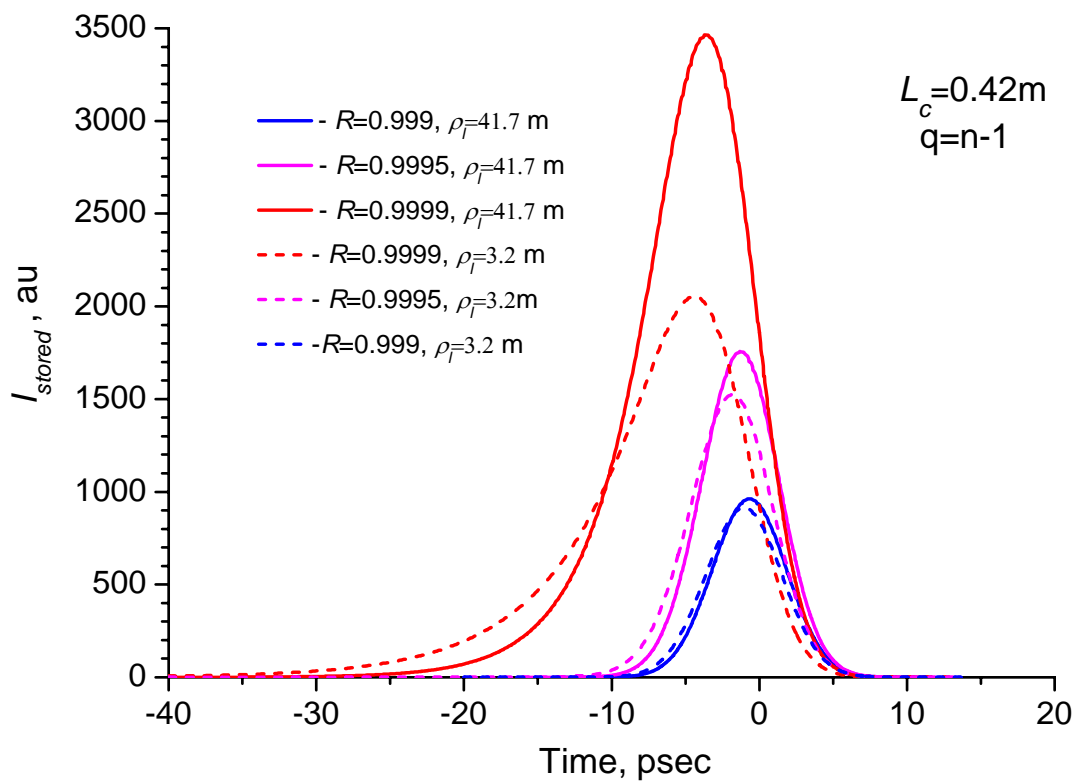

Fig. 4. Time profiles of the 7 ps laser pulse stored in 0.42m PSC with different mirror reflectivity and for two different radii of curvature of the laser mirrors.

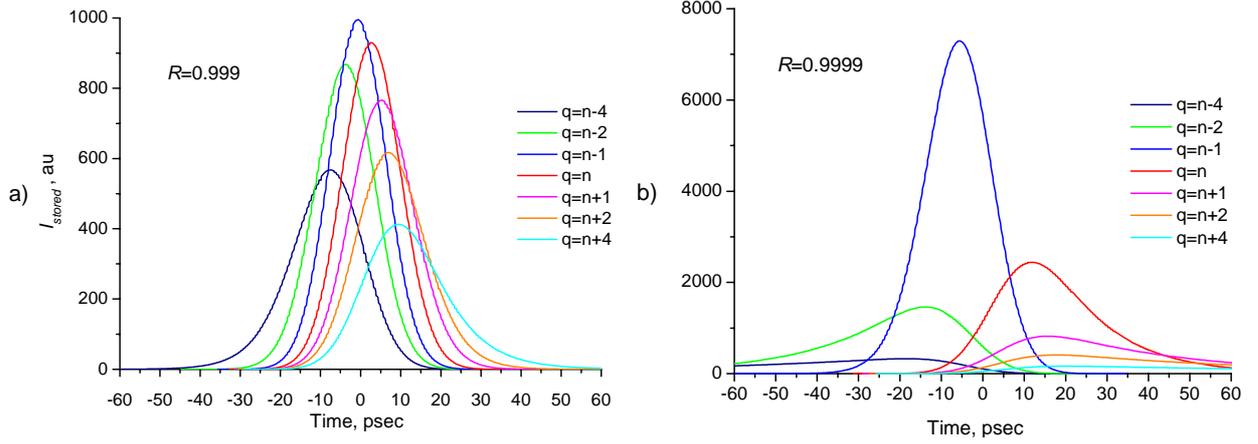

Fig. 5. Time profiles of the 20 ps laser pulse stored in 0.42m PSC with different mirror reflectivity.

We also performed calculations for the 20 ps laser pulse in order to understand how much the considered effects will be alleviated for the longer pulses. The results presented in Figure 5, when compared with those in Figure 3 (a, c), show:
- a number of PSC axial modes which can be used for pulse stacking in the medium-reflectivity version increases while in the high-reflectivity version only $q=n-1$ case is admissible;
- peak power in the high-reflectivity version essentially increases against the case of 7ps pulses, and for optimal matching $q=n-1$ it approaches the ideal matching value.

Peak power gain which is amplitude gain squared doesn't take into account pulse widening of the stored pulse. When this widening is within controllable limits it is useful to define the power gain factor $k_{ph}$ as a figure of merit that describes a pulse-stacking efficiency:

$$k_{ph} = \int_{-T_{rep}/2}^{T_{rep}/2} I_{stored}(t) \bigg/ \int_{-T_{rep}/2}^{T_{rep}/2} I_{incident}(t), \tag{21}$$

where $I_{stored}(t)$ and $I_{incident}(t)$ are power intensity of the stored and incident laser pulse, respectively. This parameter is displayed in Fig. 6 versus cavity mode shift $q-n$ for two values of PSC mirrors reflectivity and two pulse durations. One can see that the long-pulse results show a systematical excess over the short-pulse data except the case of optimal mode matching for $R=0.999$, where both points coincide giving maximal value of $10^3$ (no dispersion effects).

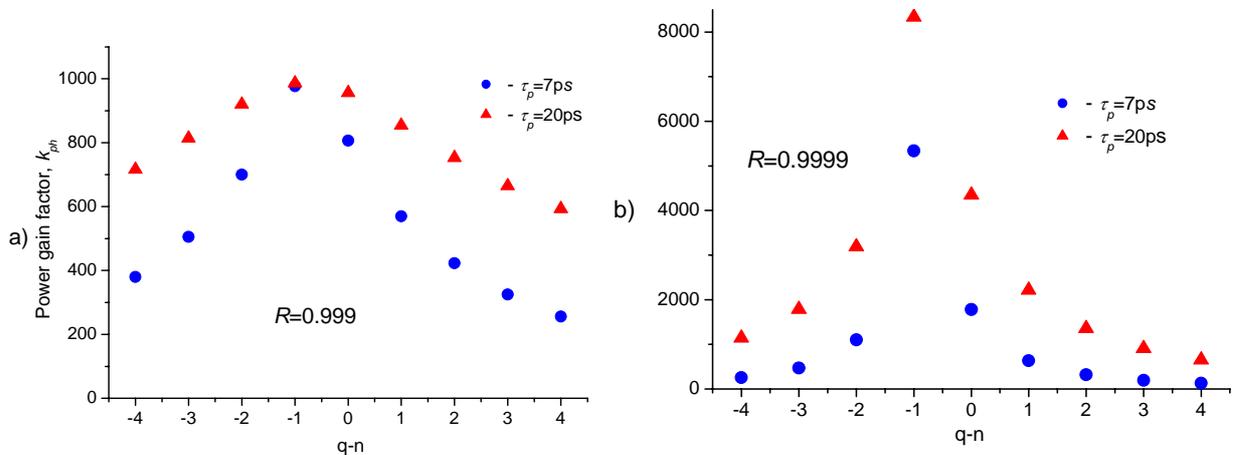

Fig. 6. Power gain factor as a function of the cavity mode shift $q-n$ for medium reflectivity mirrors (a) and high reflectivity mirrors (b)

The power gain factor $k_{ph}$ can be obtained experimentally by measuring the incident, reflected and transmitted power [8], while it is impossible to measure directly the pulse width inside the pulse-stacking cavity. One has to apply a scheme like that, discussed in [10] for extracting the stacked pulse from the PSC after desired number of round-trips, and to measure parameters of the extracted pulse.

## *Conclusion*

We have analyzed the effects of mode matching between the laser cavity and pulse-stacking cavity upon the stored pulse width and stored power. It is shown that for high reflectivity mirrors ($R \approx 0.9999$) high power gains can be obtained only if the laser carrier frequency coincides with the definite axial mode frequency of the PSC, namely with that second to the fundamental mode. For the laser pulse parameters, required for X-ray production in the compact storage ring, it means that both cavity lengths have to differ by one half of the laser wavelength, and these conditions have to be maintained during all period of generation. This task is the present state-of-the-art of laser optics technology, and it requires much efforts, both financial and scientific, to develop a laser pulse stacking system with power gains up to $10^4$.